%% file: main.tex
\documentclass[sigconf,natbib,screen]{acmart}
\pdfoutput=1

\input{packages}
\input{definitions}
\input{authors}
\input{meta}

\begin{document}

\settopmatter{printfolios=true}

\title{A Multi-Agent Conversational Recommender System}

\begin{abstract}
Due to strong capabilities in conducting fluent, multi-turn conversations with users, Large Language Models (LLMs) have the potential to further improve the performance of Conversational Recommender System (CRS).
Unlike the aimless chit-chat that LLM excels at, CRS has a clear target. So it is imperative to control the dialogue flow in the LLM to successfully recommend appropriate items to the users.
Furthermore, user feedback in CRS can assist the system in better modeling user preferences, which has been ignored by existing studies.
However, simply prompting LLM to conduct conversational recommendation cannot address the above two key challenges.

In this paper, we propose \fullmodel (\model) which contains two essential modules. 
First, we design a multi-agent act planning framework, which can control the dialogue flow based on four LLM-based agents.
This cooperative multi-agent framework will generate various candidate responses based on different dialogue acts and then choose the most appropriate response as the system response, which can help \model plan suitable dialogue acts.
Second, we propose a user feedback-aware reflection mechanism which leverages user feedback to reason errors made in previous turns to adjust the dialogue act planning, and higher-level user information from implicit semantics.
We conduct extensive experiments based on user simulator to demonstrate the effectiveness of \model in recommendation and user preferences collection.
Experimental results illustrate that \model demonstrates an improvement in user interaction experience compared to directly using LLMs.
\end{abstract}

\ccsdesc[500]{Information systems~Recommender system}
\ccsdesc[500]{Information systems~Users and interactive retrieval}

\keywords{Conversational recommendation, Large language models, Multi-agent, In-context learning}

\maketitle
\input{intro.tex}
\input{related_work}

\input{method}

\input{experiment}
\input{conclusion}


\bibliographystyle{ACM-Reference-Format}

\end{document}

%% file: packages.tex
\usepackage{tcolorbox}
\usepackage{booktabs} 
\usepackage{algorithm, algpseudocode}
\usepackage{amsmath}
\usepackage{graphics}
\usepackage{epsfig}
\usepackage{graphicx}
\usepackage{subfigure}
\usepackage{balance}
\usepackage{multirow}
\usepackage{mathrsfs}
\usepackage{acronym}
\usepackage{placeins}
\usepackage[inline]{enumitem}
\usepackage{fancyhdr}
\usepackage{bm}
\usepackage{tabularx}
\usepackage{ragged2e} 
\usepackage{makecell}
\usepackage{graphicx}
\usepackage{makecell}
\usepackage{colortbl}
\usepackage{xcolor}

\usepackage{xspace}
\usepackage{epstopdf}

%% file: definitions.tex
\AtBeginDocument{%
  \providecommand\BibTeX{{%
    \normalfont B\kern-0.5em{\scshape i\kern-0.25em b}\kern-0.8em\TeX}}}

\newcommand{\aka}{\emph{a.k.a.,}\xspace}
\newcommand{\eg}{\emph{e.g.,}\xspace}

\newcommand{\ignore}[1]{}

\newcommand{\model}{MACRS\xspace}
\newcommand{\fullmodel}{Multi-Agent Conversational Recommender System\xspace}

\newcommand{\sbbkgrnd}{\cellcolor{gray!35}}

%% file: authors.tex
\author{Jiabao Fang}
\authornote{Both authors contributed equally to this research.}
\email{202215111@mail.sdu.edu.cn}
\affiliation{
\institution{Shandong University}
\country{China}
}

\author{Shen Gao}
\authornotemark[1]
\email{shengao@sdu.edu.cn}
\affiliation{
\institution{Shandong University}
\country{China}
}

\author{Pengjie Ren}
\email{jay.ren@outlook.com}
\affiliation{
\institution{Shandong University}
\country{China}
}

\author{Xiuying Chen}
\email{xiuying.chen@kaust.edu.sa}
\affiliation{
\institution{KAUST}
\country{Kingdom of Saudi Arabia}
}

\author{Suzan Verberne}
\email{s.verberne@liacs.leidenuniv.nl}
\affiliation{
\institution{Leiden University}
\country{Netherland}
}

\author{Zhaochun Ren}
\authornote{Corresponding author.}
\email{z.ren@liacs.leidenuniv.nl}
\affiliation{
\institution{Leiden University}
\country{Netherland}
}

%% file: meta.tex
\copyrightyear{2024}
\acmYear{2024}
\setcopyright{rightsretained}
\acmConference[SIGIR '24]{Proceedings of the 47th International ACM SIGIR Conference on Research and Development in Information Retrieval}{July 14--18, 2024}{Washington D.C., USA}
\acmBooktitle{Proceedings of the 47th International ACM SIGIR Conference on Research and Development in Information Retrieval (SIGIR '24), July 14--18, 2024, Washington D.C., USA}
\acmISBN{}

%% file: intro.tex
\section{Introduction}
Unlike traditional recommender systems which solely rely on users' historical behaviors, conversational recommender system (CRS) engages in natural language interactions with the users, allowing the users to better express their needs~\cite{jannach2021survey}.
Existing CRS frameworks are generally categorized into two streams:
\begin{enumerate*}[label=(\arabic*)]
    \item The attribute-based CRS learns about user preferences by asking questions about the item attributes and generate responses based on the template. And the users only need to reply ``yes/no'' to the questions~\cite{lei2020interactive, xu2021adapting}. This approach lacks flexibility as users cannot actively pose questions to the system, while the system can only reply through human-annotated templates.
    \item The generation-based CRS focuses on generating human-like responses, where users can express their intention without limitation~\cite{zhou2020improving, wang2022towards, ren2022variational, zhang2023variational}.
Most previous generation-based methods are based on either pre-defined knowledge-graphs or smaller-scale generative models, which result in poor generalizability and infeasibility in real-world scenarios. 
\end{enumerate*}

Due to powerful language generation ability and the extensive parameterized knowledge embedded in Large Language Models (LLMs), CRS with an LLM base has recently received more and more attention~\cite{friedman2023leveraging}. 
Existing CRS with LLMs generally follows two paradigms: 
(1) \textit{LLM with external RecSys}: In these methods, the LLM is only used for response generation while an external recommender system is employed to recommend as an individual module~\cite{gao2023chat, feng2023large, liu2023conversational}.
(2) \textit{LLM-only}: The LLM in these methods not only facilitates conversations but also directly generates recommendation results given user preferences~\cite{he2023large, wang2023rethinking}. 
However, in recommendation and dialogue tasks, the LLM integrated with an external RecSys CRS relies on distinct modules (\aka recommendation module and dialogue module). 
Unfortunately, a majority of current approaches overlook the exchange of user information between these modules, consequently having an information gap between them.~\cite{wang2022towards, wang2022barcor}.
In contrast, LLM-only CRS can share useful information between the recommendation task and the dialogue task.

LLM-only CRS needs to fulfill both the recommendation task and the dialogue task, thus it typically contains various dialogue acts:
(1) An asking act that can actively elicit user preferences; 
(2) A recommending act that can provide items to users in natural language form; 
(3) A chit-chatting act that can attract user interest and guide the topic of the conversation. 
Since a conversational recommendation generally has a clearly defined target -- the goal is to recommend a specific item to a user -- an engaging CRS with LLM should be able to control the dialogue flow, meaning choose the proper dialogue act at each turn.
For example, when user preferences are ambiguous, the CRS can elicit user preferences through the asking act.
And the CRS can utilize the chit-chatting act to explore various topics and maintain an engaging conversation.
However, previous work shows that the ability of the LLM to control the target-directed dialogue flow remains limited~\cite{hong2023zero}. 
For that reason, simply prompting an LLM as a CRS can lead to unreasonable dialogue flows.
For some complex tasks that a single LLM-based agent may struggle to accomplish, researchers propose breaking them down into multiple simpler sub-tasks~\cite{qian2023communicative, qiao2024autoact} and then having multiple LLM-based agents collaborate to jointly complete the complex task. 
Similarly, CRS is also a type of complex task involving various intricate dialogue acts. 
Inspired by these multi-agent approaches~\cite{qian2023communicative, qiao2024autoact, hong2023metagpt}, we propose to employ a multi-agent framework on the CRS. 
However, designing the architecture of these agents, defining their interactions, and collaborative methods to enhance the performance of the CRS remains challenging.

Moreover, the characteristic of CRS is that users and the system can engage in multi-turn interactions.
During this process, users can provide their feedback according to the system responses.
Users can actively clarify their preferences in their feedback.
And this feedback also encompasses their satisfaction towards the recommended items.
Therefore, this informative feedback not only allows the system to obtain more user information but also reflects the reason why the recommendation failed before.
We believe this feedback can be used to dynamically optimize the system during the interaction, but this aspect has been overlooked in previous research.

\begin{figure}[ht]
\centering
  \includegraphics[width=0.95\columnwidth]{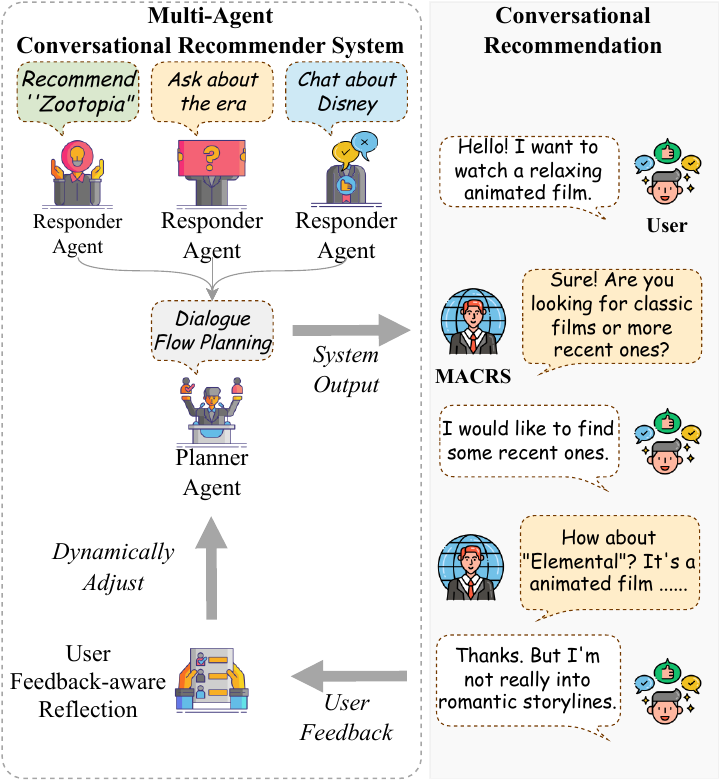}
  \caption{Example of modeling a Conversational Recommendation System (CRS) using a multi-agent framework. The right side shows a dialogue between CRS and user, while the left side depicts the internal framework of \model. The responder agent and planner agent collaboratively generate appropriate responses, while the reflection mechanism provides feedback and refined guidance to these agents, optimizing their responses to better satisfy user needs.}
  \label{fig:intro_overall}
\end{figure}
To address the above issues, we propose \fullmodel (\model), an LLM-only CRS that can efficiently plan and dynamically refine its dialogue and recommendations. 
\model comprises two essential modules (Figure~\ref{fig:intro_overall}). 
The first is \textbf{multi-agent act planning}, a cooperative LLM-based multi-agent framework which includes three responder agents and one planner agent.
In order to build an engaging dialogue flow, this framework will plan the dialogue acts for each turn.
Each responder agent in the framework is responsible for performing a dialogue act and generating a candidate response based on its dialogue act.
After collecting various candidate responses, the planner agent will comprehensively consider and select the most appropriate response as the system response.
The second module is \textbf{user feedback-aware reflection}, a dynamic optimization mechanism based on the LLM.
This mechanism leverages the user utterances as feedback to adjust the responder agents and the planner agent.
The user feedback-aware reflection contains two levels.
As user feedback in each interaction holds significant information about user preferences, including browsing history and current demand, information-level reflection becomes an essential aspect to consider.
And the information-level reflection will refine the user feedback to create higher-level information, referred to as user profiles.
These user profiles will help the responder agents generate more personalized responses in subsequent interactions.
On the other hand, when the recommendation fails, the act plan of \model needs to be adjusted in time according to the user feedback. 
Therefore, reflection will also operate on the strategy-level, which will deduce the reasons for the recommend failure and then provide suggestions and experiences to the agents.
Extensive experiments conducted on Movielens demonstrate the effectiveness of \model in recommendation and user preferences collection.

Our contributions of this work are as follows:
\begin{enumerate*}[label=(\roman*)]
\item We propose \model, an LLM-only CRS that focuses on controlling the dialogue flow for multi-turn interactions.
\item We design a multi-agent act planning framework, including three responder agents and one planner agent.
\model completes dialogue act planning by guiding these agents to collaborate.
\item We design a user feedback-aware reflection mechanism that leverages the user feedback to dynamically optimize the agent's response from the information-level and strategy-level.
\item Experimental results based on user simulator illustrate the advantage of \model in recommendation accuracy and efficiency of user preferences collection.
\end{enumerate*}

%% file: related_work.tex
\section{Related Work}
\subsection{LLM-based Autonomous Agents}
Autonomous agents are considered to be a promising step towards artificial general intelligence (AGI), which can complete tasks through self-directed planning and actions~\cite{zhou2023agents,xi2023rise}.
Meanwhile, LLMs have become powerful and fluent chatbots with convincing results in many topic domains
~\cite{zhao2023survey, min2023recent,hadi2023large}. 
Therefore, there are many attempts to use the LLM as central controllers to construct autonomous agents and thus obtain human-like decision-making capabilities~\cite{wang2023survey,park2023generative}, which are also called LLM-based agents.
For example,
~\citet{gur2023real} propose an LLM-based agent that can complete user instructions on real websites by combining canonical web actions in a program space. 
~\citet{schick2023toolformer} propose an LLM-based agent that can teach itself to use external tools via simple APIs. 

Recently, a number of researchers have started to explore LLM-based multi-agent systems, which enable the solution of more complex problems.
For example,~\citet{hong2023metagpt} propose to incorporate efficient human workflows as a meta programming approach into LLM-based multi-agent collaboration.
In the task of multi-robot cooperation,~\citet{mandi2023roco} design multiple LLM-based agents to perform high-level communication and low-level path planning for different robots. 
To the best of our knowledge, our study is the first to propose an LLM-based multi-agent system for conversation recommendation.

\subsection{Conversational Recommender Systems}
Conversational Recommender Systems (CRS) aim to recommend personalized items to users through interactive conversations~\cite{zhou2020towards}.
Conventional CRS can be broadly divided into two categories: 
The first category is attribute-based methods, in which the system clarifies user preferences by asking attribute-related questions, and users respond with ``yes/no'' to the questions.
These methods focus on when and what to ask before deciding about the item(s) to be recommended~\cite{lei2020interactive, ren2021learning}.
For example, \cite{lei2020interactive} model conversational recommendation as a path reasoning problem on a user-item-attribute graph. 
However, due to the restricted communication between the system and the user, the attribute-based methods lack flexibility.
The second category is generation-based methods, in which both users and the system can communicate using free-form natural language.
Most existing generation-based CRS rely on a knowledge graph for reasoning~\cite{zhou2020improving, chen2019towards}.
Furthermore, in order to generate more human-like responses, some prior work uses smaller-scale generative language models (e.g. DialoGPT based on GPT-2) as generators~\cite{wang2022barcor, wang2022towards}.
But due to the limited generalization ability of knowledge graphs and these smaller language models, previous generation-based CRS lack practicality in real-world scenarios.

Due to the success of LLMs, many researchers have started incorporating it into CRS. 
With its powerful natural language generation capabilities and an implicit knowledge base, LLMs have significant potential in conversational recommendation.
Existing LLM-based CRS can be broadly divided into two categories:
(1)~\citet{liu2023conversational, gao2023chat, feng2023large} incorporate the LLM with additional recommendation models. 
(2)~\citet{he2023large, wang2023rethinking} use the LLM itself as the CRS.
Although directly using LLM for CRS can share useful knowledge between different tasks, it's difficult for the LLM to control the goal-directed dialogue flow.
To the best of our knowledge, our study is the first to propose a controlling dialogue flow for the CRS completely constructed on LLM.

%% file: method.tex
\section{Methodology}

\subsection{Overview}\label{sec:overview}

Following previous CRS work~\cite{wang2022towards}, the user initiates the dialogue with the CRS by a natural language utterance to describe their demand.
Then the CRS will respond to the user with an utterance that may ask the user for more details or chit-chat with the user.
After several turns of dialogue to elicit clear user preferences, the CRS tries to recommend an item to the user.
When the user accepts the recommended item given by our CRS, we call this dialogue session a \textbf{successful sample}, and then the user will terminate the dialogue.
In our task setting, we also define a \textbf{maximum turn threshold}.
When the dialogue reaches the maximum turn threshold, but the CRS still does not provide product suggestions that the user is satisfied with, we have a fallback strategy in which the CRS will give a recommendation list containing 10 items before terminating the dialogue.
Thus, if the CRS cannot recommend an item that satisfies the user demand before reaching the maximum turn threshold and going to the fallback strategy, this dialogue session is referred to as an \textbf{unsuccessful sample}.

\begin{figure*}[ht]
\centering
  \includegraphics[width=1.85\columnwidth]{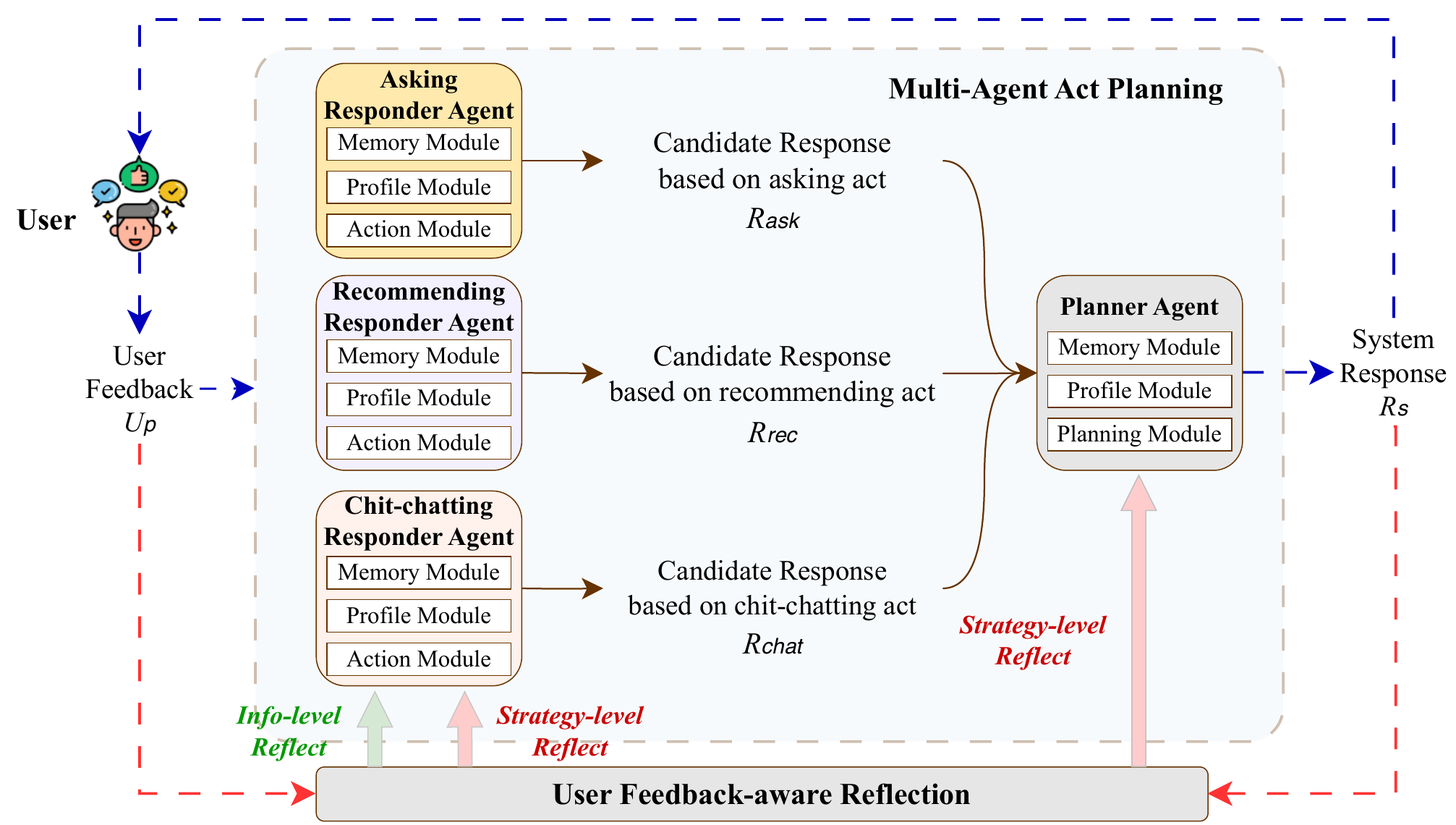}
  \caption{The overview of \model, which contains two modules: multi-agent act planning and user feedback-aware reflection. The \textit{multi-agent act planning module} generates the response according to the user feedback. The \textit{user feedback-aware reflection module} summarizes the high-level user profile and dialogue strategy suggestions which helps the responder agents to optimize their response.}
  \label{fig:overall-arch}
  \vspace{-4mm}
\end{figure*}

In this paper, to tackle this task, we introduce \fullmodel (\model), an LLM-only CRS that can flexibly control the dialogue flow and dynamically optimize the system action with user feedback.
\model consists of two modules.
The first is the multi-agent act planning, a cooperative multi-agent framework that integrates four LLM-based agents for planning the multiple dialogue acts, as shown in \S~\ref{sec:framework}.
Secondly, we propose a user feedback-aware reflection mechanism, which can leverage the user feedback to adjust the agents in \model, as shown in \S~\ref{sec:reflection}.
Figure~\ref{fig:overall-arch} shows an overview of \model.

\subsection{Multi-Agent Act Planning}\label{sec:framework}

In order to achieve an engaging interaction, a CRS generally can use various dialogue acts.
Therefore, it is essential for the CRS to effectively plan the dialogue flow by selecting suitable dialogue acts for each turn.
The dialogue flow planning plays a crucial role in enhancing the efficiency and attractiveness of the CRS.
To accomplish this complex planning task, we design a multi-agent act planning framework that contains multiple responder agents and one planner agent.
The responder agents are responsible for generating various candidate responses based on different dialogue acts.
The planner agent is responsible for devising a dialogue act plan that can develop an appropriate and fluent dialogue flow.

\subsubsection{Responder Agent}

An engaging CRS should incorporate a variety of dialogue acts during multi-turn interactions.
Therefore, we designed three types of responder agents to implement various dialogue acts including asking, chit-chatting, and recommending.
The asking responder agent $\pi^r_{ask}$ aims at asking user questions to elicit their preferences.
The recommending responder agent $\pi^r_{rec}$ shows an item to the user that it predicts the user may be interested in.
In order to improve the user experience when interacting with CRS, we cannot repeatedly recommend and ask questions but need to have timely chit-chats with users about the items using the chit-chatting responder agent $\pi^r_{chat}$. 
These chit-chats can enhance the user's interest in interacting with CRS, and can also better understand user preferences, like a real salesperson.
These responder agents share the same agent architecture which is composed of three modules: memory module, profiling module, and action module. 
All these modules are defined through in-context learning with instructions in the LLM prompt.

\textbf{The Memory module} contains the dialogue history $D_h$, the user profile $U_p$ (as mentioned in \S~\ref{sec:info-reflection}) and strategy-level suggestions $S$ (as mentioned in \S~\ref{sec:stra-reflection}). 
The dialogue history $D_h$ comprises the user utterances and system responses from previous turns.
The user profile $U_p$ contains the user's current demand and browsing history.
The strategy-level suggestions $S$ are the guidance for the responder agents in the current turn and are learned from user feedback in previous turns.

\textbf{The Profiling module} refers to the instructions and indicates the role of the responder agent.
Obviously, responder agents for different dialogue acts need to complete different task objectives, we separately design the instructions for each of the agents:


(1) Instruction $I_{ask}$ for asking responder agent $\pi^r_{ask}$:


\begin{tcolorbox}[colback=black!1!white,colframe=black!57!white]
You should elicit user preferences by asking questions.
\end{tcolorbox}

(2) Instruction $I_{chat}$ for chit-chatting agent $\pi^r_{chat}$:


\begin{tcolorbox}[colback=black!1!white,colframe=black!57!white]
You should chit-chat with the user to learn about their preferences.
You can express your admiration for certain item elements to guide the conversation towards them, thereby gaining insights into the user preferences regarding those elements.
\end{tcolorbox}

(3) Instruction $I_{rec}$ for recommending agent $\pi^r_{rec}$:

\begin{tcolorbox}[colback=black!1!white,colframe=black!57!white]
You should recommend an item to user and generate an engaging description about the item.
\end{tcolorbox}

\textbf{The Action module} generates three candidate responses $R_{ask}$, $R_{chat}$ and $R_{rec}$ based on the memory and profile:

\begin{equation}
\begin{gathered}
R_{*} = \pi_r^{*}(I_{*}, D_h, U_p, S_{*}),
\end{gathered}
\end{equation}
where $*$ indicates``ask'', ``rec'' or ``chat'' respectively.

\subsubsection{Planner Agent}

In order to control the dialogue flow, it is important to select the most suitable response from the candidate responses $R_{ask}, R_{chat},$ and $R_{rec}$ as the system response $R_s$.
To tackle this task, the planner agent $\pi_{p}$ utilizes multi-step reasoning to select responses across multiple dimensions, such as informativeness and engagement.
For example, when the available user preferences are insufficient, the planner agent reasons which candidate response can potentially yield a higher user information gain.
Then the planner agent will select the candidate response $R_{ask}$ of the ask responder agent as the final response to the user to gather more user preferences.
To achieve this goal, we propose a novel architecture of planner agent $\pi_{p}$ which is composed of a memory module, a profiling module, and a planning module:

\textbf{The Memory module} contains the dialogue history $D_h$, the dialogue act history $A_h$, and corrective experiences $E$ (introduced in \S~\ref{sec:stra-reflection}). 
The dialogue act history $A_h$ records the dialogue act selected by the planner agent in previous dialogue turns.
The corrective experiences $E$ are learned from user feedback in past turns.

\textbf{The Profiling module} indicates both its role and the background knowledge.
Instruction $I_{plan}$ for the planner agent $\pi_{p}$:
\begin{tcolorbox}[colback=black!1!white,colframe=black!57!white]
Choose one of the candidate responses based on three different dialogue acts. These three dialogue acts are: recommending, asking, and chit-chatting.
\end{tcolorbox}

\textbf{The Planning module} performs multi-step reasoning based on its memory and its profile to make an appropriate dialogue act selection.

$\bullet$ \textit{Step 1:} Review past acts in the dialogue act history $A_h$ and avoid the repetition of the same act across multiple turns.

$\bullet$ \textit{Step 2:} Determine whether the available user preferences are sufficient based on the dialogue history $D_h$. If so, the recommending act is a suitable choice.

$\bullet$ \textit{Step 3:} If the available user preferences are insufficient, determine which candidate response is more appropriate or can yield greater user information gain.

After multi-step reasoning, the planner agent $\pi_{p}$ will output a selection, which in turn determines the final system response $R_s$:
\begin{equation}
\begin{gathered}
R_s = \pi_p(I_{plan},R_{ask},R_{rec},R_{chat},D_h,A_h,E).
\end{gathered}
\end{equation}

\subsection{User Feedback-aware Reflection}\label{sec:reflection}

\begin{figure}[ht]
\centering
  \includegraphics[width=\columnwidth]{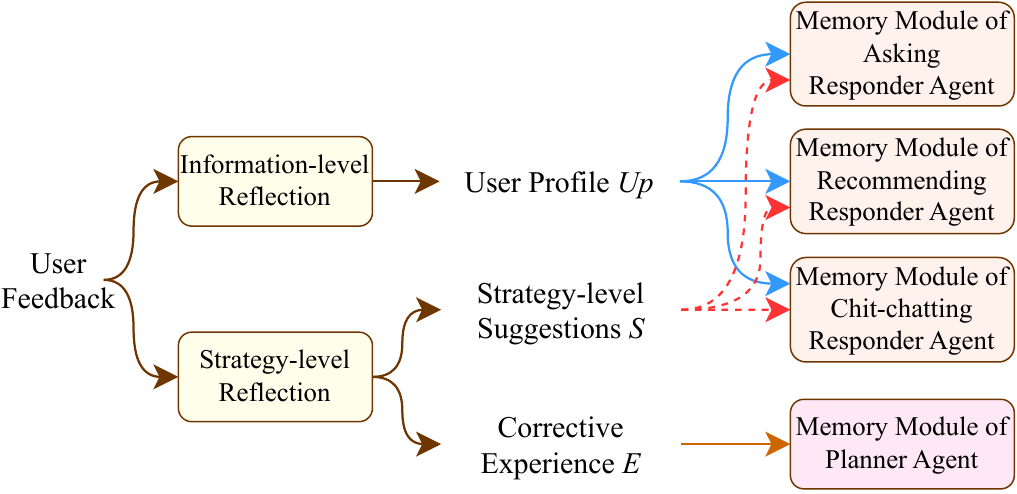}
  \caption{Overview of user feedback-aware reflection mechanism.}
  \label{fig:reflect}
\end{figure}

During the interactions with the CRS, the user provides the feedback utterance $U_f$ to the system response. 
On the one hand, users will actively express their preferences in the feedback by providing more detailed descriptions or responding to questions.
On the other hand, users will also implicitly convey their satisfaction with the recommendations and occasionally clarify the reasons behind their satisfaction.
In order to leverage this valuable feedback $U_f$ to dynamically optimize our agents, we propose a user feedback-aware reflection mechanism that operates at two levels. 
Figure~\ref{fig:reflect} provides a illustration for the proposed mechanism.
The first information-level reflection $\pi_{\text{info\_reflect}}$ infers user preferences from user feedback and summarizes them as user profiles $U_p$.
The second strategy-level reflection $\pi_{\text{stra\_reflect}}$ adjusts the dialogue act plan by generating suggestions $S$ and corrective experiences $E$ based on the user's clarifications to the recommendations.
Again, all strategies are implemented through in-context learning.

\subsubsection{Information-level Reflection}\label{sec:info-reflection}

We propose the information-level reflection mechanism $\pi_{\text{info\_reflect}}$ which summarizes user feedback to higher-level information, referred as to user profiles $U_p$.
The user profiles $U_p$ contain current user demand and browsing history: (1) The current user demand is stored in a dictionary format, comprising multiple key-value pairs. 
Each key-value pair represents an item attribute that the user is interested in. 
And these key-value pairs are extracted from the user feedback by using this mechanism $\pi_{\text{info\_reflect}}$. (2) The browsing history includes multiple items mentioned by the user, which are extracted from the user feedback $U_f$.
It helps the responder agents effectively capture user preferences and also reminds them to avoid generating duplicate content.

In dialogue turn $t$, the information-level reflection mechanism $\pi_{\text{info\_reflect}}$ generates the user profile $U^{t}_p$ based on the user profile $U^{t-1}_p$ from the previous turn $t-1$, system response $R_s^{t-1}$ and user feedback $U^{t-1}_f$:
\begin{gather}
U^{t}_p = \pi_{\text{info\_reflect}}(U^{t-1}_p, R_s^{t-1}, U^{t-1}_f),
\end{gather}
where the information-level reflection mechanism $\pi_{\text{info\_reflect}}$ is implemented by prompting an LLM-based agent using the following instructions:
\begin{tcolorbox}[colback=black!1!white,colframe=black!57!white]
Please infer user preferences based on the conversation.
And combine them with the past preferences to summarize a more {complete user preferences}.
\end{tcolorbox}

After information-level reflection, the user profiles $U_p$ will be transmitted to the memory module of each responder agent.

\subsubsection{Strategy-level Reflection}\label{sec:stra-reflection}

User feedback implicitly indicates their satisfaction with the recommendations.
When a user is dissatisfied, it indicates that \model has issues in controlling the dialogue flow in the past few turns.
So we propose the strategy-level reflection mechanism $\pi_{\text{stra\_reflect}}$ which mainly focuses on dynamically adjusting the dialogue act plan when recommendation fails. 
This mechanism $\pi_{\text{stra\_reflect}}$ is implemented by prompting an LLM-based agent. 

Our proposed strategy-level reflection mechanism $\pi_{\text{stra\_reflect}}$ will first collect multi-turn trajectory $M$, which consists of three parts: user profile $U_p$ (as mentioned in \S~\ref{sec:info-reflection}), system response $R_s$ and user feedback $U_f$. 
Then the trajectory $M$ from turn $i$ to turn $t$ can be represented as:
\begin{gather}
M^i_t = \left( (U_p^{i}, R_s^{i}, U_f^{i}), \dots, (U_p^{t}, R_s^{t}, U_f^{t})\right).
\end{gather}
When the system provide a failed recommendation in turn $t$, this mechanism $\pi_{\text{stra\_reflect}}$ reflects on the trajectory from turn $i$ to turn $t$, where turn $i$ refers to the turn immediately following the last recommendation failure.
If turn $t$ is the first recommendation failure, then $i = 1$.
To reflect the mistakes the system has made in the trajectory $M^i_t$, strategy-level reflection generates an error summary $E_s$:
\begin{gather}
E_s = \pi_{\text{stra\_reflect}}(M^i_t),
\end{gather}
where the error summary $E_s$ is a descriptive natural language representation that encompasses errors made by both the responder agents and the planner agent.
The instruction of error summary $E_s$ generation is as follow:

\begin{tcolorbox}[colback=black!1!white,colframe=black!57!white]
Based on your past action trajectory, your goal is to write a few sentences to explain why your recommendation failed as indicated by the user utterance.
\end{tcolorbox}

To avoid repeating the same mistake, the strategy-level reflection mechanism $\pi_{\text{stra\_reflect}}$ adjusts the agents by generating an error summary $E_s$.
There are two main parts in the error summary $E_s$: the strategy-level suggestions $S$ and the corrective experiences $E$.
The strategy-level suggestions $S$ indicate the dialogue strategies that each agent needs to complete in the next dialogue turn and will be used in the memory module of each responder agent (\eg ask, rec and chat).
For example, a user rejects the recommendation and responds with ``I want to watch more classic films'', which is vague.
The mechanism can suggest the asking responder agent $\pi^r_{ask}$ to elicit specific preferences regarding the release time of films.
Secondly, the strategy-level reflection summarizes the suggestions $S$ into corrective experiences $E$ and stores them in the memory module of the planner agent.
These experiences will influence its subsequent planning process.
The instruction of strategy-level suggestions $S$ and corrective experiences $E$ is as follow:
\begin{tcolorbox}[colback=black!1!white,colframe=black!57!white]
You need to generate several suggestions to ``Recommending Agent'', ``Asking Agent'' and ``Chit-chatting Agent''. Then you should report the suggestions to the ``Planning Agent'' as experiences.
\end{tcolorbox}


%% file: experiment.tex
\input{main_result}

\section{Experimental Setup}

We evaluate the effectiveness of \model in both user preferences collection and recommendation.

\subsection{User Simulator}

To evaluate the performance of \model in multi-turn interactions, we develop a LLM-based user simulator inspired by~\cite{wang2023rethinking}.
We construct the user simulator based on ChatGPT and define the preference of the user simulator using the browsing history and target item information.
In prior work~\cite{wang2023rethinking}, the target item is directly given to the user simulator, which leads to the simulator describing the target item in excessive detail.
We argue that this setting does not align with real-life scenarios, in which users usually have fuzzy preferences~\cite{li2023eliciting}.
In order to create a more realistic and natural user simulator, we use ChatGPT to summarize the information of the target item into multiple keywords, referred to as target item profiles.
Then we replace the name of target items with the target item profiles and request the user simulator to avoid overly direct descriptions.
The instruction prompt of the user simulator is:
\begin{tcolorbox}[colback=black!1!white,colframe=black!57!white]
You are a user chatting with an assistant for movie recommendation in turn. 
Your browsing history can reflect your past preferences. And you will seek recommendations from the assistant based on the \colorbox{yellow!30!white}{[target movie information]}.
\end{tcolorbox}
\noindent where the \colorbox{yellow!30!white}{[target movie information]} represents the target item profiles.

\subsection{Implementation Details}

In our experiments, all \model-C variants and the ChatGPT baseline use the gpt-3.5-turbo-0613 version, and the \model-L variant and Llama2 baseline use the Llama-2-70b-chat-hf version which is open-sourced.
We use the temperature 0.0 in our experiments.
And we update the memory modules of agents in \model after each turn, meaning that only the suggestions and experiences from the previous turn are retained.
For a fair comparison, we maintain consistency in the item summary and user utterance of the first turn across all experiments.
We set the maximum threshold of turns as 5. 
And when the CRS is still unsuccessful in the 5th turn, it will directly generate a recommendation list in the 6th turn.

\subsection{Evaluation Metrics}

To quantitatively measure the performance of \model, we design three automatic metrics:

(1) \textbf{Success Rate}: 
To evaluate the overall efficiency of CRSs, we calculate the rate of successful samples (definition can be found at \S~\ref{sec:overview}) over all the test sets:
\begin{gather}
\text{Success\_Rate} = \frac{N_{su}}{N},
\end{gather}
where $N$ is the number of test set samples, and $N_{su}$ is the number of successful samples.

(2) \textbf{Hit Ratio@K}:
Since our method has a fallback strategy of recommending an item list at the end of the conversation, we also include this part of the recommendation results in the final recommendation effect evaluation.
This allows us to evaluate the efficiency of information gathering.
We calculate the number of top-$K$ hits in recommendation lists, where $K=5, 10$. 
We also add the number of successful samples in the final calculation:
\begin{gather}
\text{Hit\_Ratio} = \frac{\sum_{i=1}^{N_{un}}hit(i) + N_{su}}{N},
\end{gather}
where $N_{un}$ is the number of unsuccessful samples (definition can be found at \S~\ref{sec:overview}). 
And when the target item appears in the recommendation list, $hit(i) = 1$. 
Otherwise, $hit(i) = 0$.

(3) \textbf{Average Turns}: 
To evaluate the recommendation efficiency of the CRS, we calculate the average number of turns in the conversation:
\begin{gather}
\text{Average\_Turns} = \frac{\sum_{i=1}^N NT_i}{N},
\end{gather}
where $NT_i$ is the number of turns in sample $i$.

\subsection{Dataset}

We conduct our experiments on the MovieLens~\cite{10.1145/2827872} which is a benchmark dataset for movie recommendation. 
It contains 162,541 users, 62,423 movies and 25,000,095 ratings. 
To construct the user simulator, we truncate the interaction sequence to 20 items, utilizing them to create the browsing history.
The 21st item is then used as the prediction target item.
Then we filter the browsing history and only retain the items that belong to the same category as the target item.
We randomly select 100 samples and ensured that the length of their browsing history is not less than 5.

\subsection{Comparison Methods}

To evaluate the effectiveness of \model, we compare it to four strong CRS baseline methods:

\noindent (1) \textbf{\texttt{KBRD}}~\cite{chen2019towards} contains a recommendation module and a Transformer-based dialogue module. It leverages an external knowledge graph to enhance the semantics of entities mentioned in the dialogue history. 

\noindent (2) \textbf{\texttt{BARCOR}}~\cite{wang2022barcor} uses the bidirectional encoder of BART~\cite{lewis2020bart} as the recommendation module and its auto-regressive decoder as the dialogue module.


\noindent (3) \textbf{\texttt{ChatGPT}}~\footnote{\url{https://chat.openai.com/}} is a closed-source LLM from OpenAI. We use the version gpt-3.5-turbo-0613. We construct a powerful single-agent CRS based on it using the same prompt as our model.

\noindent (4) \textbf{\texttt{Llama2}~\footnote{\url{https://huggingface.co/meta-llama/Llama-2-70b-chat-hf}}} is an open-source LLM that has been trained using both supervised fine-tuning and reinforcement learning with human feedback~\cite{touvron2023llama}.
We use the \texttt{Llama2} model with 70 billion parameters.
We construct it as a single-agent method to plan the dialogue flow and compare it with our multi-agent framework.

We propose two variants of \model: \textbf{\model-C} and \textbf{\model-L}, which use the \texttt{ChatGPT} and \texttt{Llama2} as the backbone respectively. To demonstrate the effectiveness of each module in our proposed \model framework, we employ several ablation models in our experiments:

\noindent (1) \textbf{\texttt{\model w/o SR}}: We remove the \textbf{s}trategy \textbf{r}eflection (SR) module in \model, which can verify the effectiveness of the reflection on the dialogue strategy.

\noindent (2) \textbf{\texttt{\model w/o IR}}: We remove the \textbf{i}nformation \textbf{r}eflection (IR) module in our user feedback-aware reflection which generates the user profile, which verifies the effectiveness of high-level user preference information.

\noindent (3) \textbf{\texttt{\model w/o SR+IR}}: We remove both of the \textbf{s}trategy \textbf{r}eflection and \textbf{i}nformation \textbf{r}eflection (SR+IR) to verify the effectiveness of the reflection mechanism in \model.

All the ablation studies are conducted on \textbf{\model-C}.

\begin{figure*}[th]
\centering
  \includegraphics[width=1.8\columnwidth]{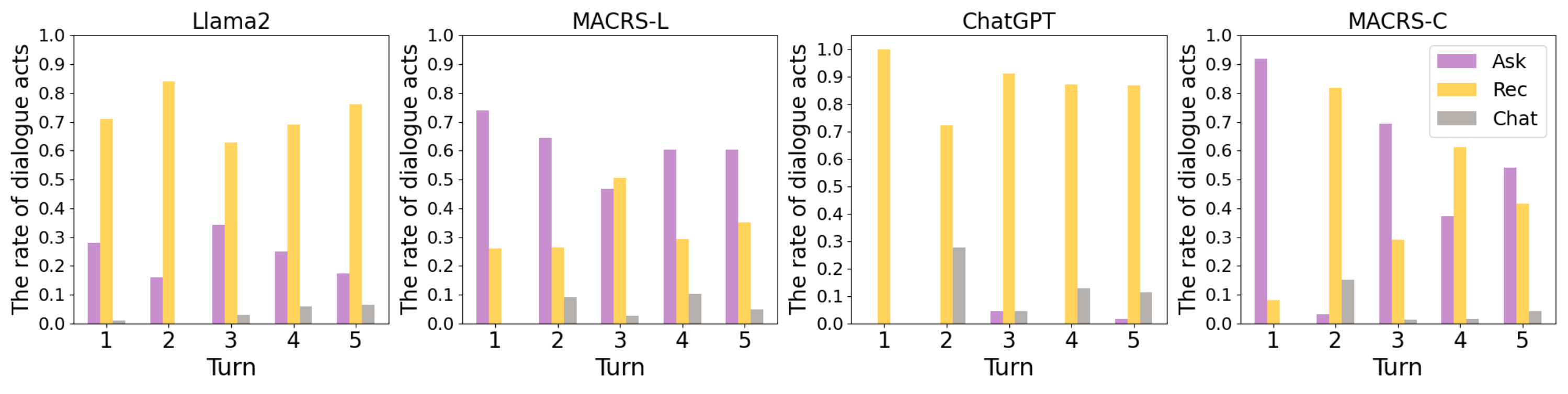}
  \caption{The ratio of different dialogue acts chosen in each turn of dialogue. The diversity in dialogue act selection of our proposed \model is remarkably enhanced compared to the corresponding backbone model, and it enhances the user experience.}
  \label{fig:acts_rate}
  \vspace{-3mm}
\end{figure*}

\section{Experimental results}

\subsection{Main Results}

Table~\ref{table:main_result} shows the performance of our proposed \model and baselines in terms of four metrics.
We see that the LLM-only CRSs (\eg \texttt{ChatGPT} and \texttt{Llama2}) demonstrate superior performance compared to traditional CRS methods (\eg \texttt{BARCOR} and \texttt{KBRD}). 
On the one hand, this is because \texttt{BARCOR} and \texttt{KBRD} rely on knowledge graphs and face limitations in successfully recommending items that are not present in the knowledge graph.
On the other hand, they rely on smaller-scale generative models that exhibit weaker reasoning and natural language understanding capabilities than LLMs.

We also find that \model-C outperforms its backbone LLM \texttt{ChatGPT} in terms of all metrics.
This suggests that with the help of our proposed \model framework, the LLM-only CRS can improve in both user preference gathering and recommendation accuracy.
Although the \model-L outperforms the \texttt{Llama2} in terms of all the recommendation accuracy metrics (\eg Success Rate and Hit Ratio), we observe that \model-L has a slightly higher average number of turns compared to the \texttt{Llama2}. 
We suspect that this is because \model method recommends more cautiously, which gives it an advantage in both success rate and recommendation accuracy.
In contrast, the single-agent methods (\eg \texttt{ChatGPT} and \texttt{Llama2}) tend to directly and frequently recommend items to the user, and this behavior will lead to higher efficiency.
However, it is obvious that such frequent recommendations not only risk user dissatisfaction but are also inefficient for capturing user preferences. 
A more detailed analysis can be found in \S~\ref{sec:act_statistics}.

Moreover, while \texttt{Llama2} exhibits enhanced performance compared to \texttt{ChatGPT} as a single agent, \model-L lags behind \model-C across all metrics. 
We speculate the reason for this phenomenon is the \texttt{Llama2}'s weaker ability to follow the complex instructions than \texttt{ChatGPT}, which results in suboptimal outcomes within our proposed multi-agent framework.

\begin{figure}[htb]
\centering
  \includegraphics[width=0.85\columnwidth]{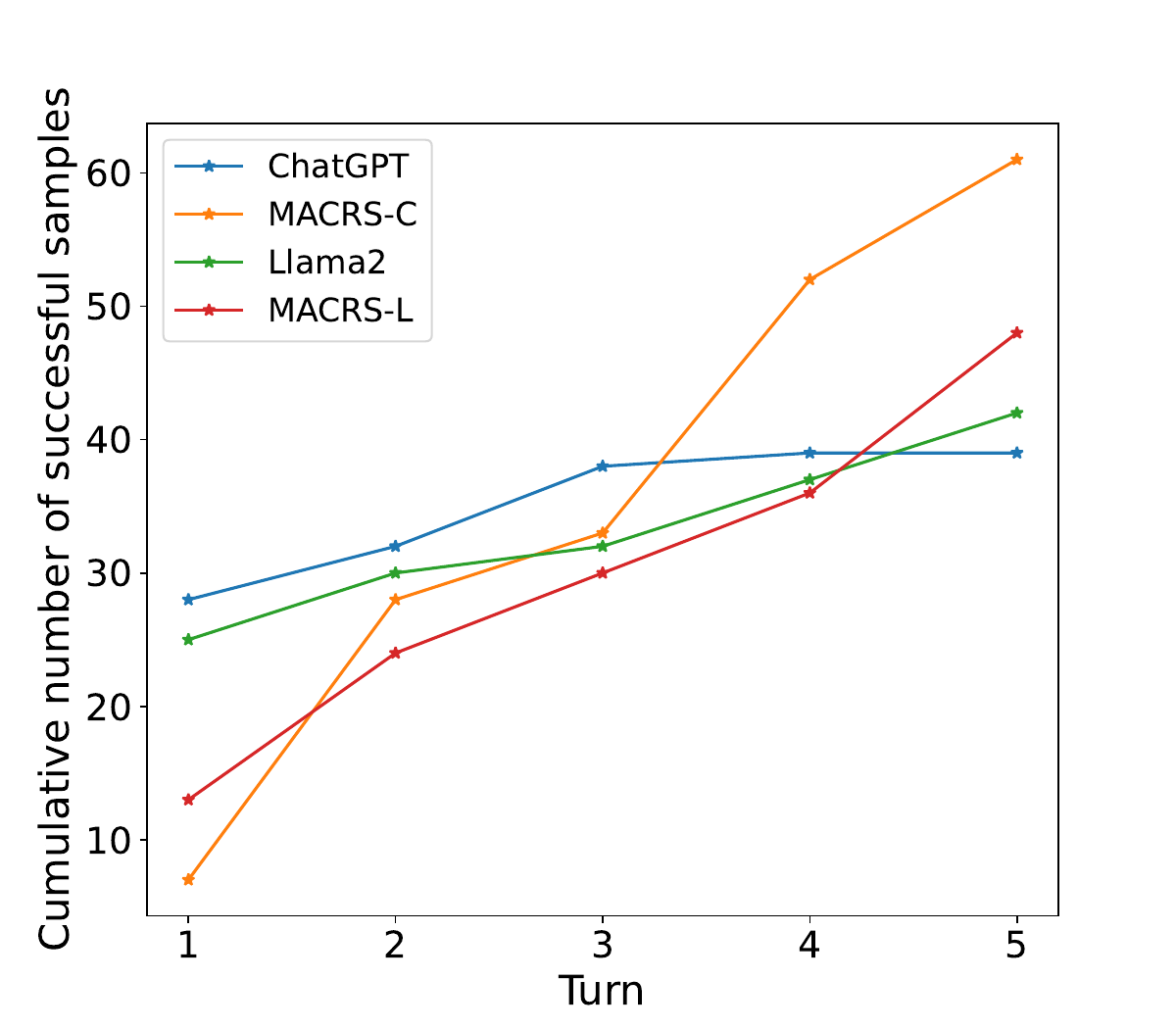}
  \caption{Cumulative number of successful samples for each dialogue turn.}
  \label{fig:turn_success}
  \vspace{-4mm}
\end{figure}

\subsection{Ablation Study} 

To evaluate the effectiveness of each module in \model, we conduct ablation studies with three variant models: (1) only using the multi-agent act planning without both levels of reflection (\aka \model w/o SR+IR); (2) only using the information-level user feedback-aware reflection (\aka \model w/o SR); (3) only using the strategy-level user feedback-aware reflection (\aka \model w/o IR).
The performance of ablation models is shown in Table~\ref{table:main_result}, and we can find that all of the ablation models perform less promising than the best model \model-C, which demonstrates the preeminence of each module in \model.
We can also find that the \model w/o SR performs worse than the \model w/o IR, which demonstrates that strategy-level suggestion and corrective experience are more important of the CRS.

\begin{figure}[bht]
\centering
  \includegraphics[width=\columnwidth]{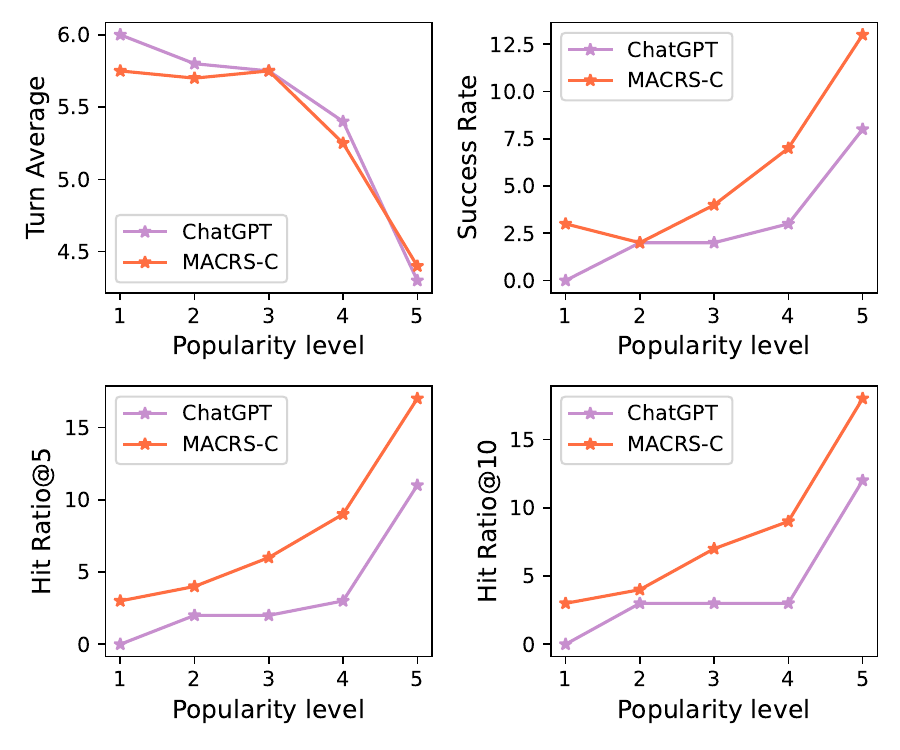}
  \caption{Performance of different models with respect to the popularity level of item.}
  \label{fig:popularity}
  \vspace{-4mm}
\end{figure}

\subsection{Discussion}

\subsubsection{Analysis of different dialogue acts ratio}\label{sec:act_statistics}

In order to compare the differences in dialogue act planning between \model and single-agent CRS, we conduct a statistical analysis comparing the dialogue act selection in each turn.
As shown in Figure~\ref{fig:acts_rate}, our method has better diversity in choosing dialogue acts than other baseline models. 
Other baseline methods (\eg \texttt{ChatGPT}) tend to directly recommend or ask the user's preference in each round of dialogue. 
This will lead to a poor user experience with the CRS, as users are often impatient with being constantly recommended boring items.
In addition, Figure~\ref{fig:acts_rate} shows that our method tends to ask the user's preference in the first round of dialogue, make a trial recommendation in the second round, and then adjust the dialogue act based on the previous interaction in the subsequent turns. 
\model also inserts a small amount of chit-chat acts at appropriate times during the dialogue to enhance the user experience and increase the appeal of the CRS to users.
We expect this conversational behavior of the CRS to be more natural and human-like and will therefore be preferred by users.

We also conduct a statistical analysis on the number of cumulative successful samples at each turn $T$, where $T = \{1, 2, 3, 4, 5\}$.
As shown in Figure~\ref{fig:turn_success}, since \model tends to ask about user preferences in the initial turns, while \texttt{ChatGPT} tends to directly make recommendations, \texttt{ChatGPT} has more accumulative successful examples when $T = \{1, 2\}$. 
However, as \texttt{ChatGPT} lacks explicit modeling of user preferences, the cumulative count of successfully recommended samples gradually decreases in later interaction turns. 
This experiment also demonstrates the importance of actively obtaining user preference information.


\subsubsection{Analysis of the influence of item popularity}

Popularity bias, a well-known phenomenon in recommendation systems~\cite{abdollahpouri2020connection, wei2021model, lin2022quantifying}, can lead to reduced user satisfaction when the target item is unpopular.
We conducted a statistical analysis to examine the performance differences of \model under varying levels of target item popularity and we compared it with the LLM-based CRS baseline \texttt{ChatGPT}.
In the experiments, we quantify the popularity of an item based on the number of interactions and categorize it into five levels: top 10\%, top 10\%-20\%, top 20\%-30\%, top 30\%-50\%, and bottom 50\%. 
For each level, we randomly selected 20 samples.

The results are shown in Figure~\ref{fig:popularity}.
The experimental results indicate that \model outperforms the baseline in all popularity levels of items.
Moreover, \model demonstrates remarkable recommendation accuracy when dealing with challenging samples characterized by lower item popularity. 
This can be attributed to the effective dialogue act planning in \model, which enables it to gather more comprehensive user preference information.

On the contrary, the single-agent method \texttt{ChatGPT} tends to make more frequent recommending acts, which can help improve the overall efficiency and reduce the number of average number of dialogue turns on the test set. 
However, this approach has a negative effect on low-popularity items, which can make it difficult for these CRS methods to recommend low-popularity items to users, as they lack a detailed understanding of user preferences.

\begin{table}[!t]
\centering
\caption{The influence of strategy-level suggestions.}
\label{table:suggest_eval}
\begin{tabular}{@{}ccc @{}}
\toprule
\textbf{\makecell[c]{Response \\ w/ Suggestions Wins}} & \textbf{Tie} & \textbf{\makecell[c]{Response \\ w/o Suggestions Wins}} \\

\midrule

$45.4\%$  & $33.0\%$  & $21.6\%$ \\

\bottomrule
\end{tabular}
\vspace{-4mm}
\end{table}

\subsubsection{The influence of strategy-level suggestions}

In order to evaluate the strategy-level suggestions generated by the strategy-level reflection mechanism, we compared the responses with suggestions and the responses without suggestions. 
We use ChatGPT~\footnote{https://chat.openai.com/} as the evaluator. We instruct the evaluator to compare the responses according to multiple criteria including (1) recommendation accuracy, (2) user information gain, (3) conciseness, and (4) engagement.
Then the evaluator selects the superior response based on these criteria.
The result shown in Table~\ref{table:suggest_eval} indicates that the strategy-level suggestions can enhance the quality of the responses.

\subsection{Case Study}

\begin{figure}[bht]
\centering
  \includegraphics[width=\columnwidth]{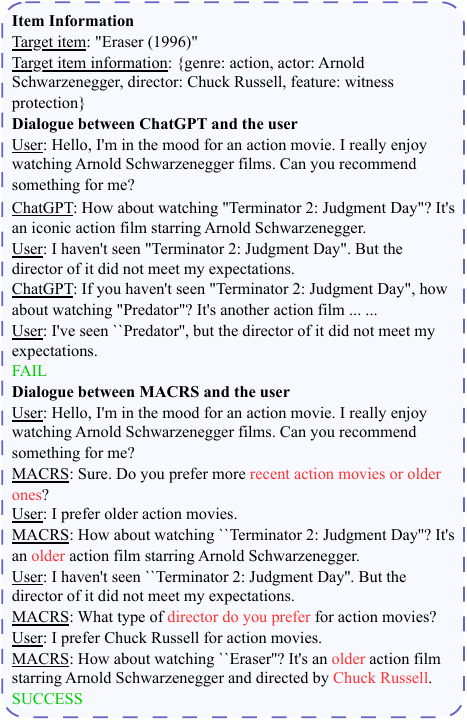}
  \caption{Case study for a conversational recommendation.}
  \label{fig:good_case}
  \vspace{-4mm}
\end{figure}

In Figure~\ref{fig:good_case}, we show an example of the dialogue between the user and \model and between the user and the baseline method.
The text in \textcolor{green}{green} indicates whether the user accepts the CRS recommended item, the text in \textcolor{red}{red} represents dialogue acts that are beneficial for recommendation. 
From this case, we can find that the \texttt{ChatGPT} fails to recommend the proper item to the user since it does not further explore user preferences in the 2nd turn.
When the user chats with \model, \model accurately identifies user implicit preferences for directors when the user replies with ``the director of it did not meet my expectations''.
It is essential for the CRS to detect implicit preferences from user feedback and further explore them to uncover more explicit preferences.
This observation demonstrates that our proposed user feedback-aware reflection module can help the CRS adjust the dialogue act plan based on user feedback. 

%% file: main_result.tex
\begin{table*}[htb]
\centering
\caption{Conversational recommendation performance comparison. $\ddagger$ indicates significant improvement over \texttt{ChatGPT} with $p \le 0.01$ according to a Student's t test.}
\label{table:main_result}

\begin{tabular}{lcccc}
\toprule
\textbf{Method} & Success Rate ($\uparrow$) & Average Turns ($\downarrow$) & Hit Ratio@5 ($\uparrow$) & Hit Ratio@10 ($\uparrow$) \\

\midrule
\multicolumn{4}{@{}l}{\emph{Traditional CRS Baselines}}\\

\texttt{KBRD}~\citep{chen2019towards}
& 0.00 & 6.0 & 0.00 & 0.01
\\

\texttt{BARCOR}~\citep{wang2022barcor}
& 0.03 & 5.9 & 0.03 & 0.07
\\

\midrule
\multicolumn{4}{@{}l}{\emph{LLM-based CRS Baselines}}\\
\texttt{Llama2}
& 0.42 & 4.34 & 0.47 & 0.51
\\

\texttt{ChatGPT}
& 0.39 & 4.24 & 0.44 & 0.51
\\

\midrule

\multicolumn{4}{@{}l}{\emph{Our Proposed Method}}\\
\model-L
& 0.48 & 4.49 & 0.55 & 0.60
\\

\model-C
& \textbf{0.61}$^{\ddagger}$ & \textbf{4.19}$^{\dagger}$ & \textbf{0.77}$^{\ddagger}$ & \textbf{0.80}$^{\ddagger}$ 
\\
\sbbkgrnd \model-C w/o IR    & \sbbkgrnd 0.53 & \sbbkgrnd 4.42 & \sbbkgrnd 0.72 & \sbbkgrnd 0.77 \\
\sbbkgrnd \model-C w/o SR    & \sbbkgrnd 0.52 & \sbbkgrnd 4.42 & \sbbkgrnd 0.67 & \sbbkgrnd 0.73 \\
\sbbkgrnd \model-C w/o SR+IR & \sbbkgrnd 0.51 &\sbbkgrnd 4.40  & \sbbkgrnd 0.64 & \sbbkgrnd 0.66 \\ 
\bottomrule
\end{tabular}
\end{table*}

%% file: conclusion.tex
\section{Conclusion}

In this paper, we introduced \fullmodel (\model), an LLM-only multi-agent Conversational Recommender System (CRS) that leverages several different agents to tackle the dialogue flow planning and integrate the user feedback to optimize the dialogue planning dynamically.
We first propose a novel multi-agent dialogue planning module to separately model each dialogue act using LLM-based responder agents and schedule these responder agents according to the dialogue strategy and user feedback by using the planner agent.
Then we propose a user feedback-aware reflection module that conducts the reflection on both information-level and strategy-level to produce high-level user preference descriptions and dialogue strategy suggestions for the subsequent dialogue.
These two modules can collaborate to optimize the dialogue flow and increase the user experience and recommendation accuracy.
Experiments conducted on a benchmark CRS dataset demonstrate that our proposed \model can boost the performance of recommendations significantly and have a better and more engaging user experience than with the existing CRS methods.